\documentclass[twocolumn,showpacs,preprintnumbers,amsmath,amssymb,superscript address, revsymb,prb]{revtex4}

\usepackage[dvips]{graphicx}
\usepackage{mathptmx}
\usepackage{dcolumn}

\newcommand{\etal}{\emph{et al.}}
\newcommand{\defect}[1]{$\{I,{\rm H}_{#1}\}$}
\newcolumntype{.}{D{.}{\cdot}{3.10}}
\newcommand{\dtwo}[2]{$\{{\rm #1}_{\rm #2}\}$}
\newcommand{\bdtwo}[2]{${\bf\{{#1}_{#2}\}}$}
\newcommand{\dfour}[4]{$\{{\rm #1}_{\rm #2},{\rm #3}_{\rm #4}\}$}
\newcommand{\bdfour}[4]{${\bf\{{#1}_{#2},{#3}_{#4}\}}$}
\newcommand{\dsix}[6]{$\{{\rm #1}_{\rm #2},{\rm #3}_{\rm #4},{\rm #5}_{\rm #6}\}$}
\newcommand{\bdsix}[6]{${\bf\{{#1}_{#2},{#3}_{#4},{#5}_{#6}\}}$}

\begin{document}

\title{Hydrogen/nitrogen/oxygen defect complexes in silicon from
computational searches}


\author{Andrew J. Morris \footnote{Email: ajm255@cam.ac.uk.}}
\affiliation{Theory of Condensed Matter Group, Cavendish Laboratory, University of Cambridge, J. J. Thomson Avenue, Cambridge CB3 0HE, United Kingdom}

\author{Chris J. Pickard}
\affiliation{Department of Physics and Astronomy, University College London, Gower St, London WC1E 6BT, United Kingdom}

\author{R. J. Needs}
\affiliation{Theory of Condensed Matter Group, Cavendish Laboratory, University of Cambridge, J. J. Thomson Avenue, Cambridge CB3 0HE, United Kingdom}

\date{\today{}}

\begin{abstract} 
Point defect complexes in crystalline silicon composed of hydrogen,
nitrogen, and oxygen atoms are studied within density-functional
theory (DFT).  \emph{Ab initio} Random Structure Searching (AIRSS) is
used to find low-energy defect structures.
We find new lowest-energy structures for several defects: the
triple-oxygen defect, $\{{\rm 3O}_{\rm i}\}$, triple oxygen with a nitrogen
atom, $\{{\rm N}_{\rm i},{\rm 3O}_{\rm i}\}$, triple nitrogen with an oxygen atom,
$\{{\rm 3N}_{\rm i},{\rm O}_{\rm i}\}$, double hydrogen and an oxygen atom,
$\{{\rm 2H}_{\rm i},{\rm O}_{\rm i}\}$, double hydrogen and oxygen atoms,
$\{{\rm 2H}_{\rm i},{\rm 2O}_{\rm i}\}$ and four hydrogen/nitrogen/oxygen complexes,
$\{{\rm H}_{\rm i},{\rm N}_{\rm i},{\rm O}_{\rm i}\}$, $\{{\rm 2H}_{\rm i},{\rm N}_{\rm i},{\rm O}_{\rm i}\}$,
$\{{\rm H}_{\rm i},{\rm 2N}_{\rm i},{\rm O}_{\rm i}\}$, and $\{{\rm H}_{\rm i},{\rm N}_{\rm i},{\rm 2O}_{\rm i}\}$.  We find that
some defects form analogous structures when an oxygen atom is replaced
by a NH group, for example, $\{{\rm H}_{\rm i},{\rm N}_{\rm i},{\rm 2O}_{\rm i}\}$ and $\{{\rm 3O}_{\rm i}\}$,
and $\{{\rm H}_{\rm i},{\rm N}_{\rm i}\}$ and $\{{\rm O}_{\rm i}\}$.  We compare defect formation
energies obtained using different oxygen chemical potentials and
investigate the relative abundances of the defects.
\end{abstract}
\maketitle

\section{Introduction} 

Hydrogen (H) is a common impurity in silicon (Si) which is
particularly important as a passivator of surfaces and bulk
defects.\cite{VdWalle:ARMR:2006}
The role of hydrogen in semiconductors is highlighted in a review by
Estreicher.\cite{Estreicher:MSE:1995} Adding nitrogen (N) impurities
to silicon affects the formation of voids and may increase the
strength of the silicon crystal by immobilising dislocations, which
reduces warping during wafer processing.\cite{Sumino:JAP:1983} The
majority of silicon used in device technologies is manufactured in a
quartz crucible by the Czochralski (Cz) process, during which oxygen
(O) from the quartz readily enters the melt, see for example the
review by Newman.\cite{Newman:JPCM:2000}

A wide variety of experimental probes are used to study defects in
silicon, but it is often difficult to determine their detailed
structures from measurements alone.  Theoretical studies using
\textit{ab initio} techniques are helpful in this regard, as they are
used to calculate both the defect formation energies and some of their
experimental signatures.  For example, local vibrational modes of
defects are accessible to infra-red (IR) absorption experiments and
may also be calculated within \textit{ab initio} methods.

Throughout this paper we denote a defect complex by listing its
constituent atoms between braces, $\{A_{\rm i},B_{\rm s},\ldots\}$,
where a subscript denotes whether an atom is substitutional (s) or
interstitial (i).  For example, a defect containing two interstitial
hydrogen atoms and a substitutional nitrogen atom is denoted by
\dfour{2H}{i}{N}{s}.  Despite hydrogen and nitrogen being common
impurities in silicon, we are aware of only one \emph{ab initio}
theoretical study of their interaction.\cite{McAfee:PRB:2004} It was
concluded that both \dtwo{N}{s} and \dtwo{N}{i} defects are able to
trap hydrogen atoms, although with smaller binding energies than that
of a hydrogen atom and a vacancy.  The Fourier transform infra-red
(FTIR) vibrational line at 2967\,cm$^{-1}$ found in silicon prepared
in a hydrogen atmosphere is assigned to the N--H stretch mode of the
\dfour{H}{i}{N}{i} defect.\cite{Pajot:PRB:1999} Interstitial oxygen
may be present in Cz-Si in concentrations as high as
$10^{18}$\,cm$^{-3}$.  Oxygen impurity atoms may be used for gettering
metallic impurities, which increases the overall crystal quality, but
they also form electrically active thermal donors (TD).\cite{GADEST}
TDs can affect the local resistivity within the silicon
wafer.\cite{Harkonen:NIMPRA:2005} In addition, oxygen also interacts
with nitrogen in bulk silicon to form TDs, and small concentrations of
hydrogen can greatly enhance the formation of
TDs.\cite{Stein:JAP:1994} It has been shown that the presence of
hydrogen molecules in silicon crystals enhances the diffusion of
oxygen.  It has also been suggested that hydrogen passivates the
electrical activity of N/O complexes in Cz-Si.\cite{Pi:PSSb:2000}

It is clear from these examples that the interactions between
hydrogen, nitrogen, and oxygen impurities in silicon lead to important
and complicated behavior. Here we present the results of a
computational search for low-energy defect complexes in silicon
containing hydrogen, nitrogen, and oxygen atoms.  We confirm the
stability of many of the previously-known defects and also report the
structures and formation energies of some new low-energy defects.  To
the best of our knowledge, this is the first time that all of these
defects have been compared within a single study.

The rest of this paper is set out as follows, in Sec.\
\ref{Computational Approach} we discuss the computational methods used
in the study.  In Sec.\ \ref{formation energies} we describe the
calculation of the defect formation energies and explain our choice of
chemical potentials.  Results for H/N defects are presented in Sec.\
\ref{sec:H/N defects}, N/O defects in Sec.\ \ref{sec:N/O defects}, H/O
defects in Sec.\ \ref{sec:H/O defects}, and H/N/O defects in Sec.\
\ref{sec:H/N/O defects}.  The relative abundances of the defects are
studied in Sec.\ \ref{sec:relative abundances} and a discussion of our
results is presented in Sec.\ \ref{discussion}.

\section{Computational Approach} 
\label{Computational Approach}

\subsection{Random Structure Searching}
\label{random structure searching}

``\emph{Ab initio} random structure searching'' (AIRSS) has already
proved to be a powerful tool for finding crystalline structures of
solids under high
pressures.\cite{pickard:PRL:2006:silane,pickard:NaturePhysics:2007:hydrogen,pickard:PRB:2007:alh3,pickard:NatureMat:2008:ammonia,pickard:PRL:2009:N,pickard:PRL:2009:Li}
The basic algorithm is very simple: we take a population of random
structures and relax them to local minima in the energy.  Pickard and
Needs\cite{pickard:PRL:2006:silane} showed that ``random structure
searching'' can be applied to finding structures of point defects, and
the approach is discussed in more detail in Ref.\
\onlinecite{Morris:PRB:2008}. 

Creating the initial simulation cell is a three-stage process.  
First, we choose the size of the disruption to the perfect silicon lattice 
by defining the radius of a sphere in which the impurity atoms are to be
placed randomly. We choose radii between 3 and 7 \AA.
Secondly, we chose the number of silicon atoms to have their positions 
ramdomized.  In this study we randomize the position of 1 to 3 silicon 
atoms depending on the number of hydrogen, nitrogen and oxygen impurity atoms present. 
The randomization sphere is centered on 
the centroid of the removed silicon atoms.
Finally, the required number of hydrogen, nitrogen and oxygen impurity atoms and the chosen 1, 2 or 3
silicon atoms are placed at random positions within the sphere.
The configurations are then relaxed using the calculated
density-functional-theory (DFT) forces.

\subsection{Density-functional-theory calculations}
\label{DFT calculations}

Our calculations are performed using the Generalized Gradient
Approximation (GGA) density functional of Perdew, Burke, and Ernzerhof
(PBE).\cite{Perdew:PRL:1996} We use the plane-wave basis-set code
\textsc{castep},\cite{CASTEP:ZK:2004} with its built-in ultrasoft
pseudopotentials\cite{Vanderbilt:PRB:1990} which include non-linear
core corrections.\cite{Louie:PRB:1982} All of the results presented
here are for non-spin-polarized systems.  The multi-Baldereschi
k-point scheme described in Ref.\ \onlinecite{Morris:PRB:2008} is used
for sampling the Brillouin zone.  We perform the same computational
convergence tests as described in Ref.\ \onlinecite{Morris:PRB:2008},
finding excellent convergence with a 2$\times$2$\times$2 k-point
sampling, which we use for all the results reported here.

The searches we report in this paper are carried out in a
body-centered-cubic simulation cell which would hold 32 atoms of
crystalline silicon.  Although this is a small cell, tests indicate
that it is adequate for obtaining good structures for most of the
defects studied and reasonably accurate formation energies, while
keeping the computational cost low enough to permit extensive
searching.  In a previous study,\cite{Morris:PRB:2008} we reported
results for hydrogen/silicon complexes in silicon.  The searches were
carried out in the same 32-atom cell used here, but each of the defect
structures was re-relaxed in a 128-atom cell to obtain more accurate
structures and energies.  The defect formation energies for the
32-atom cells were not reported, but they are in fact in excellent
agreement with the 128-atom values.  Consider, for example, the
calculations for an interstitial silicon atom (denoted by $I$) and
hydrogen impurity atoms in crystalline silicon reported in Ref.\
\onlinecite{Morris:PRB:2008}.  The changes in the formation energies
on increasing the cell size from 32 to 128 atoms for the defects
\defect{}, \defect{2}, \defect{3} and \defect{4} are only $0.00$,
$-0.07$, $-0.01$ and $-0.02$\,eV, respectively.  We have also
compared our results for N/O defects in silicon with data from the
literature.  Fujita \etal\cite{Fujita:APL:2007} used DFT and 216-atom
cells to calculate the binding energies for the following defect
reactions in silicon: \dfour{2N}{i}{O}{i}+\dtwo{O}{i} $\rightarrow$
\dfour{2N}{i}{2O}{i}, \dtwo{2N}{i}+\dtwo{O}{i} $\rightarrow$
\dfour{2N}{i}{O}{i}, and \dfour{N}{i}{O}{i} + \dtwo{O}{i}
$\rightarrow$ \dfour{N}{i}{2O}{i}, obtaining 0.91, 0.98, and 1.12\,eV,
respectively, in good agreement with our 32-atom cell results of 0.80,
0.91 and 0.97\,eV, respectively.  Finally, Coutinho
\etal\cite{Coutinho:PRB:2000} used DFT and 64-atom cells to calculate
the formation energy of the \dtwo{O}{i} defect, obtaining values
between 1.989 and 1.820\,eV, depending on the basis set used, which is
in good agreement with our value of 1.90\,eV.  We should not, of
course, expect a 32-atom cell to be sufficient for all point-defect
calculations in silicon, and especially not for charged defect states,
but there is a strong case for believing it to be adequate for the
purposes of the current study.

\section{Calculating the formation energies} 
\label{formation energies}

In order to define defect formation energies it is necessary to
specify the chemical potentials of the atomic species.  The chemical
potential for silicon is taken to be the bulk silicon value, so that
the energy cost to add a silicon atom to a defect is the energy of a
silicon atom in perfect crystalline silicon.  The chemical potential
for hydrogen is derived from the lowest energy hydrogen defect in
silicon, which is an interstitial molecule at the tetrahedral site,
\dtwo{H}{2~i}, with the bond pointing along a $\left<100\right>$
direction.  This is the same chemical potential as used in our study
of hydrogen impurities in silicon.\cite{Morris:PRB:2008} We use the
most energetically favorable nitrogen defect,
\dtwo{2N}{i},\cite{Jones:PRL:1994,McAfee:PRB:2004,Fujita:APL:2005,Goss:PRB:2003,Sawada:PRB:2000}
to set the chemical potential of nitrogen.

We perform around 130 AIRSS searches with one, two, three, or four
interstitial oxygen atoms per cell.  The lowest energy \dtwo{O}{i}
defect found is the buckled-bond-centered configuration which has been
studied
extensively.\cite{Hao:PRB:2004,Yamada-Kaneta:PRB:1990,Pesola:PRB:1999,Coutinho:PRB:2000,Artacho:PRB:1997,Plans:PRB:1987}
The \dtwo{2O}{i} defect has two buckled-bond-centered oxygen atoms
bonded to a silicon atom.  This defect was reported in Ref.\
\onlinecite{Coutinho:PRB:2000} and is known as the staggered
\dtwo{2O}{i} defect.

\begin{figure}
\includegraphics*[width=6cm]{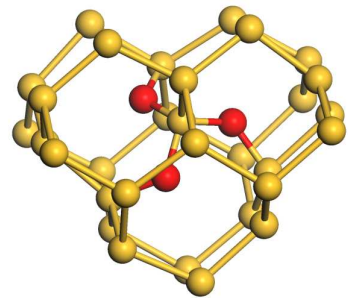}
\caption[]{(Color online) The \dtwo{3O}{i} defect is composed of three
buckled-bond-centered oxygen atoms, each bonded to a single, four-fold
coordinated silicon atom.  Silicon atoms are shown in yellow and
oxygen atoms in red.
}
\label{Fig:O3}
\end{figure}

Our searches also find the previously-reported staggered chain
\dtwo{3O}{i}
defect.\cite{Tsetseris:APL:2006,Lee:PRB:2002:LVM,Pesola:PRL:2000,Pesola:PRB:1999,Rashkeev:APL:2001,Chadi:PRL:1996}
We find a new structure, see Fig.\ \ref{Fig:O3}, with a formation
energy 0.15\,eV lower than the staggered chain.  To the best of our
knowledge this defect has not been mentioned before.  We also find the
\dtwo{4O}{i} defect which was previously reported as the O$_4$(1,1)
defect in Ref.\ \onlinecite{Lee:CPC:2001}.

Table \ref{Table:O_chem_pots} shows the formation energies per oxygen
atom of the four oxygen defects and quartz, using the chemical
potential of the \dtwo{O}{i} defect as the reference for supplying
oxygen atoms.  We note that the formation energy decreases with each
additional oxygen, showing that oxygen defects tend to aggregate, the
lowest possible energy being achieved by forming crystalline SiO$_2$
(quartz).

We consider three different chemical potentials for oxygen.  These
correspond to choosing the source of oxygen atoms as $\mu_{\{{\rm
O}_i\}}$, $\mu_{\{4{\rm O}_i\}}$, and $\mu_{\mathrm{quartz}}$.  The
chemical potential $\mu_{\{n{\rm O}_i\}}$ for oxygen with the source
of oxygen atoms being the \dtwo{{\it n}O}{i} defect is calculated as
the energy of the lowest-energy structure of the 32-atom silicon cell
containing $n$ oxygen atoms and the energy of the 32-atom bulk silicon
cell, divided by $n$.  The \dtwo{O}{i} defect gives the highest oxygen
chemical potential, while quartz gives the lowest chemical potential.
We include the chemical potential, $\mu_{\mathrm{quartz}}$, calculated
from quartz, since it is the opposite extreme to $\mu_{\{{\rm O}_i\}}$
in terms of oxygen saturation.

\begin{table}[!h]
{\centering \begin{tabular}{lr@{}l}
\hline\hline
Defect & $E_{\mathrm{f}}$ per &\ O (eV) \\
\hline
O      &  0&.00 \\
2O     & -0&.25 \\
3O     & -0&.42 \\
4O     & -0&.46 \\
Quartz & -1&.90 \\
\hline\hline
\end{tabular}\par}
\caption[]{Formation energies per oxygen atom for different oxygen
complexes, relative to the oxygen bond-centered defect
(O$_{\mathrm{bc}}$).  These values can also be thought of as the
oxygen chemical potentials relative to a source of oxygen atoms
consisting of single-oxygen defects.}
\label{Table:O_chem_pots}
\end{table}

\begin{table}[!h]
{\centering \begin{tabular}{cr@{}lr@{}lr@{}lc}
\hline\hline
Defect & &O$_{\rm i}$ & 4&O$_{\rm i}$ & Qua&rtz &Saturation\\
\hline
Bulk  & 0&.00 & 0&.00 & 0&.00 & $\surd$ \\
\hline
\dtwo{2N}{i} & 0&.00 & 0&.00 & 0&.00 & $\surd$ \\
\hline
\dtwo{O}{i}  &  0&.00 &  0&.46 & 1&.90 & $\surd$ \\
\dtwo{2O}{i} & $-$0&.50 &  0&.42 & 3&.29 &  $\surd$ \\
\bdtwo{3O}{i} & $-$1&.27 &  0&.12 & 4&.41 & $\surd$ \\
\dtwo{4O}{i} & $-$1&.85 &  0&.00 & 5&.73 & $\surd$ \\
\hline
\dtwo{H}{2~i} & 0&.00 & 0&.00 & 0&.00 & $\surd$ \\
\hline
\dfour{H}{i}{N}{i}  & 0&.45 & 0&.45 & 0&.45 & $\surd$ \\
\dfour{H}{i}{N}{s}  & 0&.81 & 0&.81 & 0&.81 & $\surd$ \\
\bdfour{2H}{i}{N}{s} & 0&.76 & 0&.76 & 0&.76 & $\times$ \\
\hline
\dfour{N}{i}{O}{i}  &  0&.43 &  0&.89 & 2&.32 & $\times$ \\
\dfour{N}{i}{2O}{i} & $-$0&.54 &  0&.39 & 3&.25 & $\times$ \\
\bdfour{N}{i}{3O}{i} & $-$0&.92 &  0&.49 & 4&.76 & $\times$ \\   
\dfour{2N}{i}{2O}{i} & $-$1&.71 & $-$0&.78 & 2&.09 & $\surd$ \\
\dfour{2N}{i}{O}{i} & $-$0&.91 & $-$0&.44 & 0&.99 & $\surd$ \\
\bdfour{3N}{i}{O}{i} &  0&.15 &  0&.61 & 2&.05 & $\times$ \\  
\hline
\dfour{H}{i}{O}{i}        &    0&.65 &    1&.11 & 2&.54 & $\times$ \\
\dfour{H}{i}{2O}{i}       & $-$0&.04 &    0&.89 & 3&.75 & $\times$ \\
\bdfour{2H}{i}{O}{i}      & $-$0&.23 &    0&.23 & 1&.67 & $\surd$ \\  
\bdfour{2H}{i}{2O}{i}     & $-$0&.76 &    0&.17 & 3&.03 & $\surd$ \\
\hline
\bdsix{H}{i}{N}{i}{O}{i}  & $-$0&.68 & $-$0&.22 & 1&.21 & $\surd$ \\ 
\bdsix{2H}{i}{N}{i}{O}{i} & $-$0&.26 &    0&.20 & 1&.63 & $\times$ \\
\bdsix{H}{i}{2N}{i}{O}{i} & $-$0&.49 & $-$0&.03 & 1&.40 & $\times$ \\
\bdsix{H}{i}{N}{i}{2O}{i} & $-$0&.85 &    0&.08 & 2&.94 & $\surd$ \\
\hline\hline
\end{tabular}\par}
\caption[]{Formation energies defined by Eq.\ (\ref{Eqn:EfNH}) in eV
for various H/N/O complexes with three different choices of the oxygen
chemical potential. The final column indicates whether the system
could form a structure with fully-saturated covalent bonding
($\surd$), or whether it cannot ($\times$), see Sec.\
\ref{discussion}.
The defect structures shown in bold have not, to the best of our
knowledge, been reported in the literature before.
}
\label{Table:defect_formation_energies}
\end{table}

The formation energy of a defect, $E_\mathrm{f}$, is defined as
\begin{equation}
E_\mathrm{f} = E_\mathrm{D} - \sum_{i} n_i \mu_i,
\label{Eqn:EfNH}
\end{equation}
where $n_i$ is the number of each atomic type $i$ in the defect cell
of energy $E_\mathrm{D}$, and the chemical potentials, $\mu_i$, are
defined above.

\begin{figure}
\includegraphics*[width=6cm]{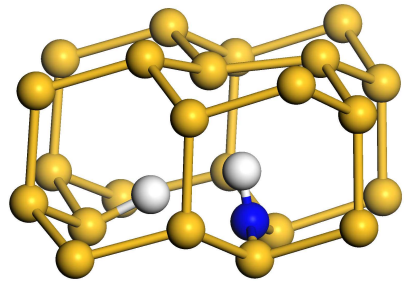}
\caption[]{(Color online) The first metastable H/N defect,
\dfour{2H}{i}{N}{s}, contains a three-fold coordinated nitrogen atom
close to a lattice site. One of the hydrogen atoms is bonded directly
to the nitrogen atom and points toward a neighboring three-fold
coordinated silicon atom.  The remaining hydrogen atom terminates a
dangling bond on another neighboring silicon atom.  The silicon atoms
are shown in yellow, the nitrogen in blue and the hydrogen atoms in
white. }
\label{Fig:2HN}
\end{figure}
\begin{figure}
\includegraphics*[width=6cm]{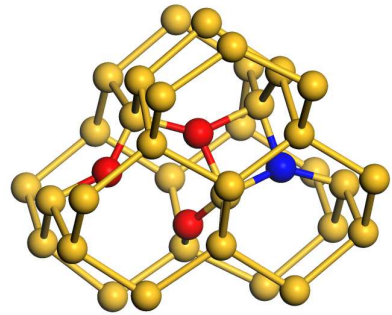}
\caption[]{(Color online) The \dfour{N}{i}{3O}{i} defect contains a
four-atom ring consisting of a nitrogen atom, an oxygen atom and two
silicon atoms.  The two other oxygen atoms are in
buckled-bond-centered positions. Note that the structure contains an
over-coordinated oxygen atom with three bonds.  The silicon atoms are
shown in yellow, the nitrogen in blue, and the oxygen atoms in red.}
\label{Fig:N3O}
\end{figure}
\begin{figure}
\includegraphics*[width=8cm]{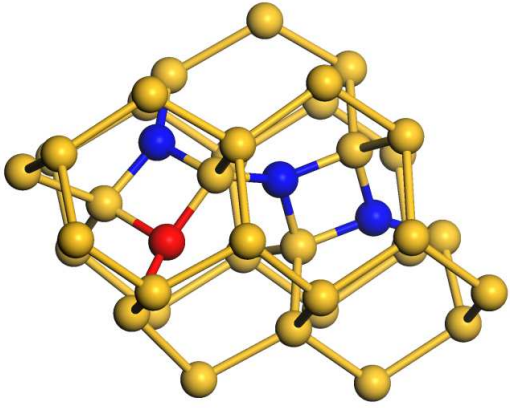}
\caption[]{(Color online) The lowest-energy \dfour{3N}{i}{O}{i} defect
consists of a four-atom ring of two nitrogen atoms and two silicon
atoms adjacent to another four-atom ring of an oxygen atom, a nitrogen
atom and two silicon atoms.  Note that the structure contains an
over-coordinated oxygen atoms with three bonds.  The silicon atoms are
in yellow, the nitrogen atoms in blue and the oxygen atom in red. }
\label{Fig:3NO}
\end{figure}
\begin{figure}
\includegraphics*[width=6cm]{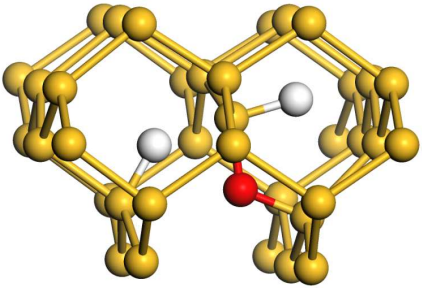}
\caption[]{(Color online) The \dfour{2H}{i}{O}{i} defect is composed
of a buckled-bond-centered oxygen atom adjacent to a distorted
\dtwo{H}{2}$^*$ defect.  The silicon atoms are shown in yellow, the
oxygen atom in red and the hydrogen atoms in white.  This defect has a
negative formation energy, showing that the metastable \dtwo{H}{2}$^*$
defect\cite{Chang:PRB:1989} is stabilized by the presence of oxygen.}
\label{Fig:2HO}
\end{figure}
\begin{figure}
\includegraphics*[width=6cm]{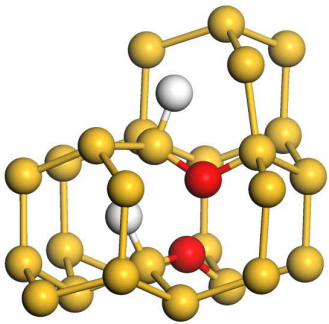}
\caption[]{(Color online) The \dfour{2H}{i}{2O}{i} defect is composed
of two adjacent \dfour{H}{i}{O}{i} defects.  The silicon atoms are
shown in yellow, the oxygen atom in red and the hydrogen atoms in
white.}
\label{Fig:2H2O}
\end{figure}
\begin{figure}
\includegraphics*[width=8cm]{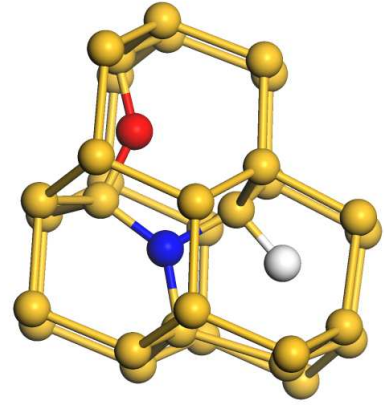}
\caption[]{(Color online) The \dsix{H}{i}{N}{i}{O}{i} defect contains
a nitrogen and silicon atom sharing a lattice site, with the dangling
silicon bond terminated by the hydrogen atom. A nearest-neighbor
silicon atom of the nitrogen atom is bonded to a buckled-bond-centered
oxygen atom.  The silicon atoms are shown in yellow, the oxygen in
red, the nitrogen in blue and the hydrogen in white. }
\label{Fig:HNO}
\end{figure}
\begin{figure}
\includegraphics*[width=8cm]{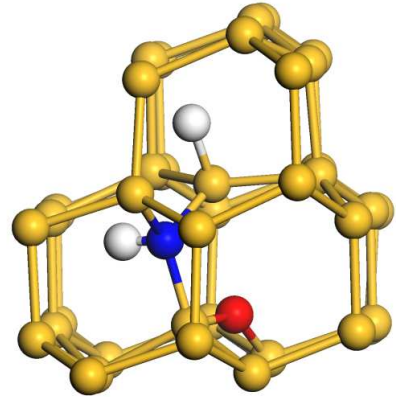}
\caption[]{(Color online) The \dsix{2H}{i}{N}{i}{O}{i} defect is
similar to the \dsix{H}{i}{N}{i}{O}{i} defect but with the nitrogen
atom acquiring the extra hydrogen atom, making it over-coordinated
with four bonds. This defect has the highest formation energy of all
the H/N/O defects studied and is therefore unlikely to form. The
silicon atoms are shown in yellow, the oxygen in red, the nitrogen in
blue and the hydrogen atoms in white.}
\label{Fig:2HNO}
\end{figure}
\begin{figure}
\includegraphics*[width=8cm]{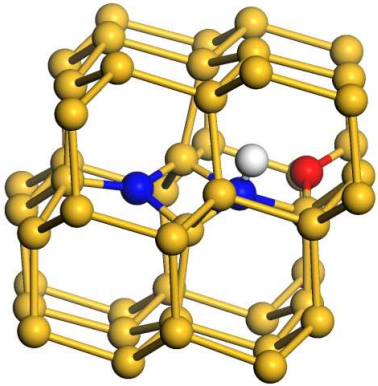}
\caption[]{(Color online) The \dsix{H}{i}{2N}{i}{O}{i} defect contains
a four-atom ring of two nitrogen atoms and two silicon atoms. One of
the nitrogen atoms is over-coordinated and is bonded to the hydrogen
atom. A nearest-neighbor silicon atom to this nitrogen atom is also
bonded to a buckled-bond-centered oxygen atom. The silicon atoms are
shown in yellow, the oxygen in red, the nitrogen in blue and the
hydrogen in white.}
\label{Fig:H2NO}
\end{figure}
\begin{figure}
\includegraphics*[width=8cm]{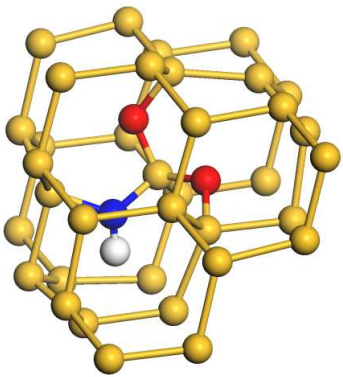}
\caption[]{(Color online) The \dsix{H}{i}{N}{i}{2O}{i} defect is very
similar to the \dtwo{3O}{i} defect, but with one of the oxygen atoms
replaced by a NH group.  The silicon atoms are shown in yellow, the
oxygen in red, the nitrogen in blue and the hydrogen in white.}
\label{Fig:HN2O}
\end{figure}

\section{H/N defects}
\label{sec:H/N defects}

We have relaxed around 250 different starting structures for the H/N
defects using cells containing 32 silicon atoms, 1 nitrogen atom and 1
hydrogen atom, and further searches with 31 silicon atoms, 1 nitrogen
atom and 1 hydrogen atom, and with 31 silicon atoms, 1 nitrogen atom
and 2 hydrogen atoms.  These are the only calculations reported in
this paper where we change the number of silicon atoms in the cell.
We find defects which we may describe as bulk silicon with an
interstitial nitrogen atom and an interstitial hydrogen atom,
\dfour{H}{i}{N}{i}, bulk silicon with a substitutional nitrogen atom
and an interstitial hydrogen atom, \dfour{H}{i}{N}{s}, and bulk
silicon with a substitutional nitrogen atom and two interstitial
hydrogen atoms, \dfour{2H}{i}{N}{s}.

The most stable H/N structure we find is the \dfour{H}{i}{N}{i} defect
with $C_s$ symmetry.  In this defect the hydrogen atom is bonded to
the buckled-bond-centered nitrogen atom, which is three-fold
coordinated, the nitrogen, hydrogen, and two silicon atoms lie in a
plane.  The N--Si bond lengths are both about 1.71\,{\AA} and the N--H
bond-length is 1.03\,{\AA}, in good agreement with McAfee
\etal,\cite{McAfee:PRB:2004} who found bond lengths of 1.73\,{\AA} and
1.05\,{\AA}, respectively.

The first metastable defect we find, \dfour{2H}{i}{N}{s}, is shown in
Fig.\ \ref{Fig:2HN} and has $C_{3v}$ symmetry.  This is quite a
complicated defect, and it is energetically slightly more favorable
than the \dfour{H}{i}{N}{s} defect which is described below.
\dfour{2H}{i}{N}{s} looks similar to the \dfour{H}{i}{N}{s} defect but
with the extra hydrogen atom bonded to the three-fold coordinated
nitrogen atom.

Finally, we report the \dfour{H}{i}{N}{s} defect.  This is similar to
the $C_{3v}$ symmetry \dtwo{N}{s} defect,\cite{McAfee:PRB:2004} but
with the hydrogen atom bonded to the three-fold coordinated silicon
atom, increasing its coordination number to four.  The three N--Si
bonds are all 1.85\,{\AA} long, the same as in \dtwo{N}{s}.  The N--H
distance is 1.95\,{\AA} and the H--Si bond is 1.80\,{\AA} long, in
good agreement with McAfee \etal\cite{McAfee:PRB:2004}

All of the H/N complexes we mention in this section have
$E_\mathrm{f}>0$, and hence they are unlikely to form spontaneously.

\section{N/O defects}
\label{sec:N/O defects}

Our searches for the N/O complexes required around 1800 random
starting structures.  These searches found all the previously-known
lowest-energy structures except the \dfour{2N}{i}{2O}{i} defect.  The
\dfour{2N}{i}{2O}{i} defect is quite large and it is probable that we
have not relaxed a sufficiently large number of starting structures to
find it.

For the \dfour{N}{i}{O}{i} defect we obtain the previously-known
interstitial ring.\cite{Ono:APE:2008} This defect has a positive
$E_\mathrm{f}$(O$_{\it i}$) of 0.43\,eV, and hence is unlikely to
form.  The structure of our \dfour{N}{i}{2O}{i} defect is the same as
found in previous
studies.\cite{Ono:APE:2008,Fujita:APL:2007,Gali:JPCM:1996} This defect
has a negative $E_\mathrm{f}$(O$_{\rm i}$) of $-0.54$\,eV, and
therefore it could form in silicon.

We are not aware of any previous reports of the \dfour{N}{i}{3O}{i}
defect, which is shown in Fig.\ \ref{Fig:N3O}. The \dfour{N}{i}{3O}{i}
defect has a negative $E_\mathrm{f}$(O$_{\rm i}$) of $-0.92$\,eV,
showing that a single bond-centered oxygen atom will bind to the
\dfour{N}{i}{2O}{i} defect.

Our searches for the \dfour{2N}{i}{2O}{i} defect did not yield the
structure reported previously,\cite{Fujita:JMS:ME:2007} which has
different oxygen positions.  We generated the structure reported by
Fujita \etal\cite{Fujita:JMS:ME:2007} ``by hand'' and found it to be
0.19\,eV lower in energy than our best structure.  The defect of
Fujita \etal\cite{Fujita:JMS:ME:2007} has a negative formation energy
of $E_\mathrm{f}$(O$_{\rm i}$) = $-1.71$\,eV, and the formation energy
remains negative even when the reference structure for the oxygen
chemical potential is taken to be that of the \dtwo{4O}{i} defect in
silicon.  It is disappointing that our searches has not given the
previously-known lowest-energy \dfour{2N}{i}{2O}{i} defect.  However,
we have included the lowest-energy known structure in our analysis of
the relative populations of the various defects presented in Sec.\
\ref{sec:relative abundances}.

We find the previously-known lowest-energy \dfour{2N}{i}{O}{i} defect,
\cite{Ono:APE:2008,Fujita:JMS:ME:2007,Gali:JPCM:1996,Jones:SST:1994}
which has a formation energy of $E_\mathrm{f}$(O$_{\rm i}$) =
$-0.91$\,eV.  This defect binds oxygen very strongly and
$E_\mathrm{f}$(O$_{\rm i}$) remains negative even with the oxygen
chemical potential from the \dtwo{4O}{i} defect.

We are not aware of any previous reports of the \dfour{3N}{i}{O}{i}
defect in the literature, which is shown in Fig.\ \ref{Fig:3NO}.  The
\dfour{3N}{i}{O}{i} defect is quite large and it may not be well
described within a 32-atom cell.

\section{H/O defects} 
\label{sec:H/O defects}

We have relaxed around 350 starting structures for the H/O defects.
The lowest energy \dfour{H}{i}{O}{i} defect that we find is composed
of a buckled-bond-centered oxygen atom adjacent to a
buckled-bond-centered hydrogen atom.  This structure was also found by
Estreicher.\cite{Estreicher:PRB:1990,Ramamoorthy:SSC:1998} This defect
has, however, been the subject of some controversy, as Jones \etal
\cite{Jones:MSF:1992,Jones:PTRSLA:1995} proposed that the hydrogen
atom is attached to a silicon atom at an anti-bonding site.  Our
search also finds this defect configuration, but we calculate it to be
0.32\,eV higher in energy than the ground state structure.

Our lowest-energy \dfour{H}{i}{O}{i} defect has a positive formation
energy, and hence it is unlikely to form.  The \dfour{2H}{i}{2O}{i}
defect is composed of two such defects in close proximity to one
another, as shown in Fig.\,\ref{Fig:2H2O}.  The defect has a negative
formation energy and, to our knowledge, it has not been presented in
the literature previously.

The lowest-energy \dfour{2H}{i}{O}{i} defect that we find (Fig.\
\ref{Fig:2HO}) adopts a \dtwo{H}{2}$^*$+O$_{\mathrm{bc}}$
configuration.  Measurements have shown an IR absorption line at
1075.1\,cm$^{-1}$ which has been assigned to a O$_{\rm i}$-H$_2$
complex.\cite{Markevich:JAP:1998} However, the
\dtwo{H}{2}$^*$+O$_{\mathrm{bc}}$ configuration of the
\dfour{2H}{i}{O}{i} defect does not contain a hydrogen molecule, see
Fig.\ \ref{Fig:2HO}.  We also find a \dfour{2H}{i}{O}{i} defect
structure containing a hydrogen molecule, but it is metastable with an
energy 0.26\,eV above our ground-state structure.  To the best of out
knowledge the \dtwo{H}{2}$^*$+O$_{\mathrm{bc}}$ defect structure has
not been reported previously.

Our most stable \dfour{H}{i}{2O}{i} defect is a
O$_{\mathrm{bc}}$+O$_{\mathrm{bc}}$+H configuration.  This defect has
also been studied by Rashkeev \etal\ (see Fig.\ 3(a) of Ref.\
\onlinecite{Rashkeev:APL:2001}), who report it to be a thermal double
donor (TDD).

The \dfour{H}{i}{2O}{i}, \dfour{2H}{i}{O}{i} and \dfour{2H}{i}{2O}{i}
defects all have negative formation energies $E_\mathrm{f}$(O$_{\rm
i}$) and are therefore likely to form in bulk silicon.

\section{H/N/O defects}
\label{sec:H/N/O defects}

We have performed around 500 structural relaxations for the H/N/O
defects.  The four defects \dsix{H}{i}{N}{i}{O}{i} (Fig.\
\ref{Fig:HNO}), \dsix{2H}{i}{N}{i}{O}{i} (Fig.\ \ref{Fig:2HNO}),
\dsix{H}{i}{2N}{i}{O}{i} (Fig.\ \ref{Fig:H2NO}), and
\dsix{H}{i}{N}{i}{2O}{i} (Fig.\ \ref{Fig:HN2O}), all have negative
formation energies when using $E_\mathrm{f}$(O$_{\rm i}$), and
\dsix{H}{i}{N}{i}{O}{i} and \dsix{H}{i}{2N}{i}{O}{i} also have
negative formation energies when using $E_\mathrm{f}$(4O$_{\rm i}$).

The \dsix{H}{i}{N}{i}{O}{i} defect has a formation energy of
$-0.68$\,eV compared with $0.43$\,eV for \dfour{N}{i}{O}{i}.  This
implies that hydrogen readily binds to the \dfour{N}{i}{O}{i} defect.
The hydrogen atom bonds to a silicon atom and breaks the third bond of
the over-coordinated oxygen atom, giving the structure shown in Fig.\
\ref{Fig:HNO}.  This results in a defect with fully saturated bonds
which is therefore quite low in energy.

The \dsix{2H}{i}{N}{i}{O}{i} defect is very similar to the
\dsix{H}{i}{N}{i}{O}{i} defect mentioned above.  The extra hydrogen
atom bonds to the nitrogen atom which is then over-coordinated, having
four bonds, see Fig.\ \ref{Fig:2HNO}.  The over-coordinated nitrogen
atom is energetically unfavorable and this defect would not readily
form from \dsix{H}{i}{N}{i}{O}{i}, however it could form from
\dfour{N}{i}{O}{i} by capturing a hydrogen molecule.

The \dsix{H}{i}{2N}{i}{O}{i} defect is an interesting case. The
nitrogen atom is over-coordinated because it binds to the hydrogen
atom, see Fig.\ \ref{Fig:H2NO}.  Since \dfour{2N}{i}{O}{i} has a
formation energy of $-0.91$\,eV and \dsix{H}{i}{2N}{i}{O}{i} has a
formation energy of $-0.49$\,eV it is unlikely that a hydrogen atom
will bind to \dfour{2N}{i}{O}{i}.

The \dsix{H}{i}{N}{i}{2O}{i} defect is very interesting.  The
fully-saturated bonding of this defect results in a low formation
energy of $-0.85$\,eV.  This is lower than the formation energy of
\dfour{N}{i}{2O}{i}, and hydrogen will therefore bind to this defect.
The \dsix{H}{i}{N}{i}{2O}{i} defect has a similar geometry to
\dtwo{3O}{i}, but with one of the oxygen atoms replaced by a HN group.
A HN group acting in a similar fashion to an oxygen atom was also
observed in the \dfour{N}{i}{H}{i} defect of
McAfee,\cite{McAfee:PRB:2004} which looks similar to the bond-centered
\dtwo{O}{i} defect.

\section{Relative abundances}
\label{sec:relative abundances}

\begin{figure}
\includegraphics*[width=8cm]{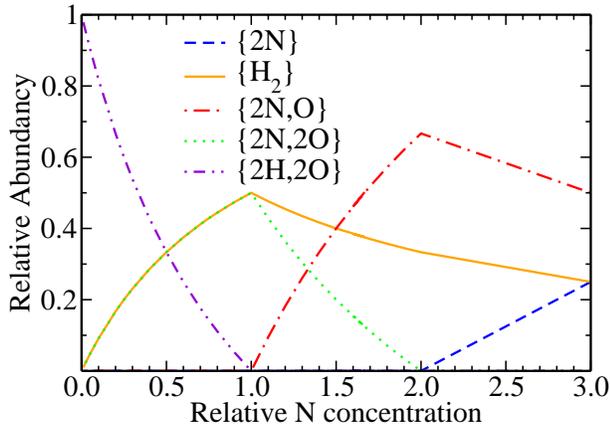}
\caption[]{(Color online) Relative concentrations of H/N/O defects in
silicon as a function of the nitrogen concentration.  At low nitrogen
concentrations (H:N:O = 1:$<$0.5:1), the predominant defect is
\dfour{2H}{i}{2O}{i}. As the nitrogen concentration increases the
concentration of \dfour{2H}{i}{2O}{i} decreases, with
\dfour{2N}{i}{2O}{i} and molecular hydrogen being formed.  As the
nitrogen concentration increases above unity, the concentration of
\dfour{2N}{i}{2O}{i} declines and \dfour{2N}{i}{O}{i} begins to form.
At nitrogen concentrations larger than two, the additional nitrogen
atoms form \dtwo{2N}{i} defects.}
\label{Fig:N_rel_conc}
\end{figure}

\begin{figure}
\includegraphics*[width=8cm]{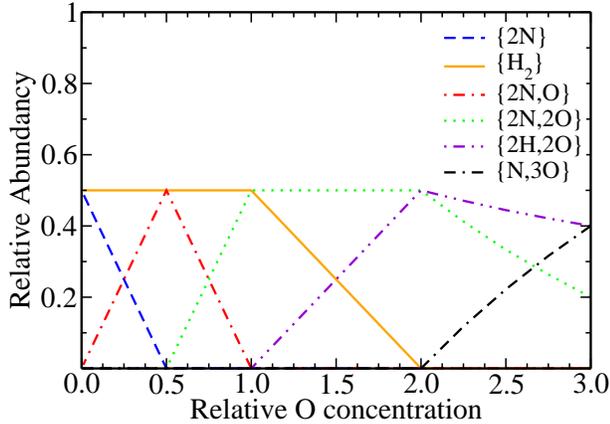}
\caption[]{(Color online) Relative concentrations of H/N/O defects in
silicon as a function of the oxygen concentration.
The behavior is complex in this case, although the general trend is
simply that the more stable defects containing larger numbers of
oxygen atoms are favored at higher oxygen concentrations.}
\label{Fig:O_rel_conc}
\end{figure}

\begin{figure}
\includegraphics*[width=8cm]{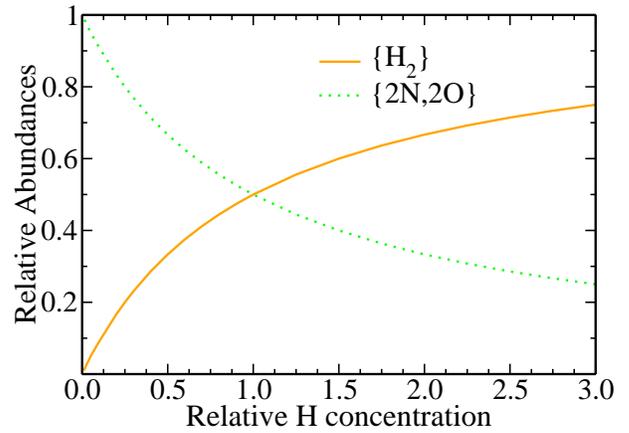}
\caption[]{(Color online) Relative concentrations of H/N/O defects in
silicon as a function of the hydrogen concentration.  Increasing the
hydrogen concentration simply generates more interstitial hydrogen
molecules.
}
\label{Fig:H_rel_conc}
\end{figure}

The formation energies of the various defects can be used to calculate
the relative abundances of the defects at zero temperature as a
function of the ratios of the H/N/O concentrations.  The relative
abundances for a chosen set of concentrations are those which minimise
the total energy, and the chemical potentials do not enter the
calculation.  However, if we allow the oxygen atoms to combine with
silicon to form quartz, it is straightforward to show that the minimum
energy is always obtained by producing as much quartz as possible, so
that no point defects containing oxygen atoms remain in the bulk, and
the hydrogen and nitrogen atoms form \dtwo{H}{2~i} and \dtwo{2N}{i}
defects.  This is inconsistent with the extensive experimental
evidence for other point defects in silicon involving hydrogen,
nitrogen, and oxygen atoms.  To gain some insight into the H/N/O
defects which may be present we must therefore limit the propensity to
form quartz and other low-energy oxygen defects in some way.  We have
chosen to present relative abundances in which the formation of
\dtwo{2O}{i}, \dtwo{3O}{i}, \dtwo{4O}{i}, up to quartz are excluded,
while allowing the formation of \dtwo{O}{i}.  This is an arbitrary
choice, but it serves to illustrate the type of behavior which may
arise.  Figs.\ \ref{Fig:N_rel_conc}, \ref{Fig:O_rel_conc}, and
\ref{Fig:H_rel_conc} show relative abundances of the defects as a
function of the H/N/O concentrations.  In each figure we keep the
concentrations of two of the species constant and equal to unity,
while the concentration of the third impurity varies from zero to
three.
The main features of Figs.\ \ref{Fig:N_rel_conc},
\ref{Fig:O_rel_conc}, and \ref{Fig:H_rel_conc} are that, for equal
concentrations of hydrogen, nitrogen, and oxygen, the dominant defects
are hydrogen molecules and the \dfour{2N}{i}{2O}{i} defect.  Varying
the nitrogen or oxygen concentration leads to the formation of a wide
variety of defects.  Increasing the hydrogen concentration, however,
only generates more hydrogen molecules.  Other scenarios can be
investigated using the data given in Table
\ref{Table:defect_formation_energies}.


\section{Discussion} 
\label{discussion}

We have presented \textit{ab initio} calculations of H/N/O complexes
in bulk silicon within DFT.  The defect complexes were generated in
32-atom supercells using the AIRSS method as outlined in Ref.\
\onlinecite{Morris:PRB:2008}.  The 32-atom cells are large enough to
give a reasonable description of most of the defects, but small enough
to allow extensive searching.  To the best of our knowledge this it
the first time that H/N/O complexes in silicon have been studied in
detail.

AIRSS searches were performed on oxygen, hydrogen, and nitrogen
defects to evaluate chemical potentials for the atoms.  Whilst
carrying out the searches on oxygen defects, we discovered a new
\dtwo{3O}{i} defect, shown in Fig.\ \ref{Fig:O3}, which is more stable
than the staggered defect which has been reported in the literature.

Our searches reproduced all of the previously-known lowest-energy H/N
complexes.  We have also described a new metastable
\dfour{2H}{i}{N}{s} defect which is more favorable than some of the
other H/N complexes presented in the literature.

The searches for N/O complexes were slightly less successful.  We were
unable to find the previously-known \dfour{2N}{i}{2O}{i} defect.  In
this case, we performed DFT calculations on the previously-known best
structure for comparison.  We discovered a new \dfour{3N}{i}{O}{i}
defect (Fig.\ \ref{Fig:3NO}) and a new \dfour{N}{i}{3O}{i} defect
(Fig.\ \ref{Fig:N3O}) which is more stable than the
\dfour{N}{i}{2O}{i} defect, implying that an oxygen atom could bind to
\dfour{N}{i}{2O}{i}.

Our searches reproduced all of the previously-known lowest-energy H/O
complexes.  We find the \dfour{H}{i}{O}{i} defect to have a positive
formation energy, and therefore it is unlikely to form, but two
adjacent \dfour{H}{i}{O}{i} defects can form a \dfour{2H}{i}{2O}{i}
defect (Fig.\,\ref{Fig:2H2O}) which has a formation energy of
$E_\mathrm{f}$(O$_{\rm i}$)$ = -0.76$\,eV.  We also find a new
\dfour{2H}{i}{O}{i} defect which is based on the previously-reported
metastable H$_2^{*}$ defect.\cite{Chang:PRB:1989} The
\dfour{2H}{i}{O}{i} defect has $E_\mathrm{f}$(O$_{\rm i}$)$ =
-0.23$\,eV, showing that the H$_2^{*}$ defect can bind an oxygen atom.

To the best of our knowledge, no previous discussion of H/N/O
complexes in silicon has appeared in the literature.  The
\dsix{H}{i}{N}{i}{O}{i} and \dsix{H}{i}{N}{i}{2O}{i} defects are more
energetically favorable than their non-hydrogen-containing
counterparts \dfour{N}{i}{O}{i}, \dfour{N}{i}{2O}{i}, showing that
many of these defects could exist in hydrogenated forms.

Another point of interest is that H/N complexes behave in an analogous
fashion to oxygen defects.  For example, the \dfour{H}{i}{N}{i} defect
adopts a buckled-bond-centered arrangement with the hydrogen atom
terminating the dangling bond on the nitrogen atom.  The nitrogen atom
takes a similar position to the oxygen atom in the O$_{\mathrm{bc}}$
defect, however without the hydrogen atom, since oxygen forms only two
covalent bonds.  The same relationship holds between the \dtwo{3O}{i}
and \dsix{H}{i}{N}{i}{2O}{i} defects, where again the H/N group in the
latter behaves like an O$_{\mathrm{bc}}$ defect in the former.

The right-hand column of Table \ref{Table:defect_formation_energies}
indicates whether or not it is possible for the indicated set of atoms
to form a fully-saturated covalently-bonded structure.  In such a
structure each silicon atom is bonded to four other atoms, each
nitrogen to three, each oxygen to two, and each hydrogen to one.  As
each bond is shared between two atoms, a fully-saturated
covalently-bonded structure is impossible if
\begin{equation}
4N_{\rm Si} + 3N_{\rm N} + 2N_{\rm O} + N_{\rm H} = {\rm \ odd\ number},
\label{Eqn:Saturation}
\end{equation}
where $N_{\rm X}$ denotes the number of atoms of type X in the cell.
We note from the data in Table \ref{Table:defect_formation_energies}
that the lowest energy defects are normally those having completely
saturated bonds.  A similar result for hydrogen defects in silicon was
reported in Ref.\ \onlinecite{Morris:PRB:2008}.  This simple and
rather obvious rule is useful for choosing combinations of impurity
atoms which might form stable defects.

Consider a host crystal containing three types of impurity atom.  (An
example might be H/N/O impurities in silicon as studied here, but
without the possibility of silicon vacancy formation which was allowed
in our study of H/N defects.) Limiting the total number of impurity
atoms to be $\leq$4 gives 34 possible cell contents for which searches
must be performed.  Increasing the number of possible impurities to
five (while maintaining the maximum number of impurities in a cell to
be $\leq$4) gives a total of 125 possible cell contents.  The extra
impurities could be other types of atoms, but they could also be host
vacancies $V$ and interstitials $I$, although this would reduce the
number of possible cell contents to 112 because introducing a host
vacancy and interstitial is the same as not introducing either.
Searching over the ``impurities'' H/N/O/$I$/$V$ in silicon, with a
maximum number of impurities of $\leq$4 and the same computational
parameters as in the present study would cost about three times as
much as the present study.  For each set of cell contents we have in
this work performed between about 20--900 structural relaxations,
which appeared to be adequate in this case, although the required
number is expected to increase rapidly with the size of the ``hole''
in the host crystal.  In some cases it may be necessary to use larger
``holes'' to obtain all of the low-energy defects, and perhaps larger
simulation cells as well.  We have automated the search procedure and
the calculations are performed in parallel.  It appears to us that
defect structure searches of the type described in this paper with up
to, say, five types of impurity are perfectly feasible with modern
computing facilities.

\begin{acknowledgments}
We are grateful to Phil Hasnip and Michael Rutter for useful
discussions.  This work was supported by the Engineering and Physical
Sciences Research Council (EPSRC) of the United Kingdom.
Computational resources were provided by the Cambridge High
Performance Computing Service.
\end{acknowledgments}


\begin{thebibliography}{10}

\bibitem{VdWalle:ARMR:2006}
C.~G. Van~de Walle and J.~Neugebauer,
\newblock Annu. Rev. Mater. Res. {\bf 36}, 179 (2006).

\bibitem{Estreicher:MSE:1995}
S.~K. Estreicher,
\newblock Mater. Sci. Eng. {\bf R 14}, 319 (1995).

\bibitem{Sumino:JAP:1983}
K.~Sumino, I.~Yonenaga, M.~Imai, and T.~Abe,
\newblock J. App. Phys. {\bf 54}, 5016 (1983).

\bibitem{Newman:JPCM:2000}
R.~C. Newman,
\newblock J. Phys: Condens. Matter {\bf 12}, R335 (2000).

\bibitem{McAfee:PRB:2004}
J.~L. McAfee, H.~Ren, and S.~K. Estreicher,
\newblock Phys. Rev. B {\bf 69}, 165206 (2004).

\bibitem{Pajot:PRB:1999}
B.~Pajot, B.~Clerjaud, and Z.-J. Xu,
\newblock Phys. Rev. B {\bf 59}, 7500 (1999).

\bibitem{GADEST}
Proceedings of the GADEST Conferences [Solid State Phenom. {\bf 108}-{\bf 109},
  97-123 (2005)], [\emph{ibid}. {\bf 95-96}, 539-586].

\bibitem{Harkonen:NIMPRA:2005}
J.~H{\"{a}}rk{\"{o}}nen, E.~Tuovinen, P.~Luukka, E.~Tuominen, Z.~Li, A.~Ivanov, E.~Verbitskaya, V.~Eremin,  A.~Pirojenko, I.~Riihimaki and A.~Virtanen,
\newblock Nucl. Instr. and Meth. {\bf 541}, 202 (2005).

\bibitem{Stein:JAP:1994}
H.~J. Stein and S.~Hahn,
\newblock J. App. Phys. {\bf 75}, 3477 (1994).

\bibitem{Pi:PSSb:2000}
X.~Pi, D.~Yang, X.~Ma, Q.~Shui, and D.~Que,
\newblock Phys. Stat. Sol. B {\bf 221}, 641 (2000).

\bibitem{pickard:PRL:2006:silane}
C.~J. Pickard and R.~J. Needs,
\newblock Phys. Rev. Lett. {\bf 97}, 045504 (2006).

\bibitem{pickard:NaturePhysics:2007:hydrogen}
C.~J. Pickard and R.~J. Needs,
\newblock Nature Physics {\bf 3}, 473 (2007).

\bibitem{pickard:PRB:2007:alh3}
C.~J. Pickard and R.~J. Needs,
\newblock Phys. Rev. B. {\bf 76}, 144114 (2007).

\bibitem{pickard:NatureMat:2008:ammonia}
C.~J. Pickard and R.~J. Needs,
\newblock Nature Materials  (2008).

\bibitem{pickard:PRL:2009:N}
C.~J. Pickard and R.~J. Needs,
\newblock Phys. Rev. Lett. {\bf 102}, 125702 (2009).

\bibitem{pickard:PRL:2009:Li}
C.~J. Pickard and R.~J. Needs,
\newblock Phys. Rev. Lett. {\bf 102}, 146401 (2009).

\bibitem{Morris:PRB:2008}
A.~J. Morris, C.~J. Pickard, and R.~J. Needs,
\newblock Phys. Rev. B {\bf 78}, 184102 (2008).

\bibitem{Perdew:PRL:1996}
J.~P. Perdew, K.~Burke, and M.~Ernzerhof,
\newblock Phys. Rev. Lett. {\bf 77}, 3865 (1996).

\bibitem{CASTEP:ZK:2004}
S.~J.~Clark, M.~D.~Segall, C.~J.~Pickard, P.~J.~Hasnip, M.~I.~J.~Probert, K.~Refson and M.~C.~Payne,
\newblock Z. Kristallogr {\bf 220}, 567 (2005).

\bibitem{Vanderbilt:PRB:1990}
D.~Vanderbilt,
\newblock Phys. Rev. B {\bf 41}, 7892 (1990).

\bibitem{Louie:PRB:1982}
S.~G. Louie, S.~Froyen, and M.~L. Cohen,
\newblock Phys. Rev. B {\bf 26}, 1738 (1982).

\bibitem{Fujita:APL:2007}
N.~Fujita, R.~Jones, S.~\"{O}berg, and P.~R. Briddon,
\newblock App. Phys. Lett. {\bf 91}, 051914 (2007).

\bibitem{Coutinho:PRB:2000}
J.~Coutinho, R.~Jones, P.~R. Briddon, and S.~\"Oberg,
\newblock Phys. Rev. B {\bf 62}, 10824 (2000).

\bibitem{Jones:PRL:1994}
R.~Jones, S.~\"Oberg, F.~Berg~Rasmussen, and B.~Bech~Nielsen,
\newblock Phys. Rev. Lett. {\bf 72}, 1882 (1994).

\bibitem{Fujita:APL:2005}
N.~Fujita, R.~Jones, J.~P.~Goss, P.~R.~Briddon, T.~Frauenheim and S.~\"{O}berg,
\newblock App. Phys. Lett. {\bf 87}, 021902 (2005).

\bibitem{Goss:PRB:2003}
J.~P. Goss, I.~Hahn, R.~Jones, P.~R. Briddon, and S.~\"Oberg,
\newblock Phys. Rev. B {\bf 67}, 045206 (2003).

\bibitem{Sawada:PRB:2000}
H.~Sawada and K.~Kawakami,
\newblock Phys. Rev. B {\bf 62}, 1851 (2000).

\bibitem{Hao:PRB:2004}
S.~Hao, L.~Kantorovich, and G.~Davies,
\newblock Phys. Rev. B {\bf 69}, 155204 (2004).

\bibitem{Yamada-Kaneta:PRB:1990}
H.~Yamada-Kaneta, C.~Kaneta, and T.~Ogawa,
\newblock Phys. Rev. B {\bf 42}, 9650 (1990).

\bibitem{Pesola:PRB:1999}
M.~Pesola, J.~von Boehm, T.~Mattila, and R.~M. Nieminen,
\newblock Phys. Rev. B {\bf 60}, 11449 (1999).

\bibitem{Artacho:PRB:1997}
E.~Artacho {\em et~al.},
\newblock Phys. Rev. B {\bf 56}, 3820 (1997).

\bibitem{Plans:PRB:1987}
J.~Plans, G.~Diaz, E.~Martinez, and F.~Yndurain,
\newblock Phys. Rev. B {\bf 35}, 788 (1987).

\bibitem{Tsetseris:APL:2006}
L.~Tsetseris, S.~Wang, and S.~T. Pantelides,
\newblock App.Phys. Lett. {\bf 88}, 051916 (2006).

\bibitem{Lee:PRB:2002:LVM}
Y.~J. Lee, M.~Pesola, J.~von Boehm, and R.~M. Nieminen,
\newblock Phys. Rev. B {\bf 66}, 075219 (2002).

\bibitem{Pesola:PRL:2000}
M.~Pesola, Y.~Joo~Lee, J.~von Boehm, M.~Kaukonen, and R.~M. Nieminen,
\newblock Phys. Rev. Lett. {\bf 84}, 5343 (2000).

\bibitem{Rashkeev:APL:2001}
S.~N. Rashkeev, M.~D. Ventra, and S.~T. Pantelides,
\newblock App. Phys. Lett. {\bf 78}, 1571 (2001).

\bibitem{Chadi:PRL:1996}
D.~J. Chadi,
\newblock Phys. Rev. Lett. {\bf 77}, 861 (1996).

\bibitem{Lee:CPC:2001}
Y.~J. Lee and R.~M. Nieminen,
\newblock Comp. Phys. Comm. {\bf 142}, 305 (2001).

\bibitem{Chang:PRB:1989}
K.~J. Chang and D.~J. Chadi,
\newblock Phys. Rev. B {\bf 40}, 11644 (1989).

\bibitem{Ono:APE:2008}
H.~Ono,
\newblock App. Phys. Express {\bf 1}, 025001 (2008).

\bibitem{Gali:JPCM:1996}
A.~Gali, J.~Miro, P.~De\'ak, C.~P. Ewels, and R.~Jones,
\newblock J. Phys: Condens. Matter {\bf 8}, 7711 (1996).

\bibitem{Fujita:JMS:ME:2007}
N.~Fujita, R.~Jones, S.~\:{O}berg, and P.~R. Briddon,
\newblock J. Mater. Sci. Mater. Electron {\bf 18}, 683 (2007).

\bibitem{Jones:SST:1994}
R.~Jones, C.~Ewels, J.~Goss, J.~Mieo, P.~De\'ak, S.~\:Oberg and F.~{Berg Rasmussen},
\newblock Semicond. Sci. Technol. {\bf 9}, 2145 (1994).

\bibitem{Estreicher:PRB:1990}
S.~K. Estreicher,
\newblock Phys. Rev. B {\bf 41}, 9886 (1990).

\bibitem{Ramamoorthy:SSC:1998}
M.~Ramamoorthy and S.~T. Pantelides,
\newblock Solid State Commun. {\bf 106}, 243 (1998).

\bibitem{Jones:MSF:1992}
R.~Jones, S.~\"{O}berg, and A.~Umerski,
\newblock Mater. Sci. Forum {\bf 83}, 551 (1992).

\bibitem{Jones:PTRSLA:1995}
R.~Jones, W.~Jackson, A.~M. Stoneham, R.~C. Newman, and M.~Symons,
\newblock Phil. Trans. Roy. Soc. Lon. Ser. A {\bf 350}, 1693 (1995).

\bibitem{Markevich:JAP:1998}
V.~P. Markevich and M.~Suezawa,
\newblock J. Appl. Phys. {\bf 83}, 2988 (1998).

\end{thebibliography}
\end{document}